\documentclass{article}

\usepackage[english]{babel}

\usepackage[top=2cm,bottom=2cm,left=2cm,right=2cm,marginparwidth=1.75cm]{geometry}
\usepackage{amsmath}
\usepackage{graphicx}
\usepackage[colorlinks=true, allcolors=blue]{hyperref}

\newcommand{\beq}{\begin{equation}}
\newcommand{\eeq}{\end{equation}}
\newcommand{\beqa}{\begin{eqnarray}}
\newcommand{\eeqa}{\end{eqnarray}}

\def\cE{ \mathcal{E} }
\def\cH{ \mathcal{H} }
\def\cI{ \mathcal{I} }

\def\commentout#1{}
\def\NEW#1{\textcolor{black}{#1}}

\def\WHT#1{\textcolor{white}{#1}}
\def\la{ \left\langle }
\def\ra{ \right\rangle }

\newcommand{\bu}{{\bf u}} 
\newcommand{\bw}{{\bf w}} 
\newcommand{\bF}{{\bf f}}
\newcommand{\bk}{{\bf k}}
\newcommand{\bq}{{\bf q}}

\newcommand{\bx}{{\bf x}}

\newcommand{\ou}{\overline{\bu}}

\newcommand{\curl}{\nabla \times}

\title{Quasi-two-dimensional Turbulence}
\author{Alexandros Alexakis \\
 Laboratoire de Physique de l’Ecole normale supérieure, ENS, \\ Université PSL, CNRS, Sorbonne Université, Université de Paris, \\ 
 F-75005 Paris, France \\ alexakis@phys.ens.fr}

\begin{document}
\maketitle

\abstract{
Many fluid-dynamical systems met in nature are {\it quasi-two-dimensional}: they are
constrained to evolve in approximately two dimensions with little or no variation along the third direction.
This has a drastic effect in the flow evolution because the properties of three dimensional turbulence are 
fundamentally different from those of two dimensional turbulence. 
In three-dimensions energy is transferred on average towards small scales, 
while in two dimensions energy is transferred towards large scales. 
%
Quasi-two-dimensional flows thus stand in a crossroad,
with two-dimensional motions attempting to self-organize and generate large scales
while three dimensional perturbations cause disorder, disrupting any large scale organization.
%
Where  is energy transferred in such systems?  
It has been realised recently that in fact the two behaviors can coexist 
with a simultaneous transfer of energy both to large and to small scales.
How the cascade properties change 
as the variations along the third direction are suppressed has lead to discovery of different 
regimes or {\it phases} of turbulence of unexpected richness in behavior.
Here,  recent discoveries on such systems are reviewed. It is described
how the transition from three-dimensional to two-dimensional flows takes place,
the different phases of turbulence met and 
the nature of the transitions from one phase to the other.
Finally,  the implications these new discoveries have on different physical systems are discussed.

\WHT{.}

\

This work is published in: \qquad \qquad  \qquad \qquad  \qquad \qquad  \qquad \qquad  \qquad \qquad \

{\bf Reviews of Modern Plasma Physics 7.1 (2023): 31.}  \qquad \qquad \qquad \qquad

\, DOI: 10.1007/s41614-023-00134-3  \qquad \qquad  \qquad \qquad  \qquad \qquad  \qquad \qquad

}

\newpage

\section{Introduction}\label{sec1}

We live in a world of (at least) three spatial dimensions.
However, for some physical systems a two dimensional description appears to provide a suitable approximation for the dynamics involved. 
This happens when one dimension is highly compactified like in thin films or due to some other physical mechanisms, like rotation or strong magnetic fields, that prevents variations along one dimension. 
In fluid dynamics such a dimension-reduction is met in a variety of systems from molecular to astrophysical scales.
At the smallest scales, such a reduction is met in the dynamics of electrons in ultra-pure materials such as graphene that display two dimensional hydrodynamic behavior
\cite{narozhny2022hydrodynamic,aharon2022direct,bandurin2018fluidity}.
In an equally exotic case, two-dimensional flows are met in fluids of light \cite{abobaker2022inverse,ferreira2022towards}. 
Two dimensional flows have also been observed in Bose-Einstein condensates in liquid Helium \cite{seo2017observation,gauthier2019giant,johnstone2019evolution}. 
Furthermore, two-dimensional dynamics are shown to be a good approximation 
for the motion of bacteria in films or thin layers \cite{sokolov2007concentration,kurtuldu2011enhancement,wu2000particle,wei2023scaling} and for the flow in  soap films 
\cite{martin1998spectra,rivera1998turbulence,vorobieff1999soap,kellay2002two}. 
Increasing in scale, plasma flows in the presence of strong magnetic fields, such as in Tokamak  devices are also known to display close to two-dimensional dynamics  \cite{Xia_2003,fujisawa2008review}.
Finally, at the planetary scale, flows are very often constrained to two dimensional motions due to the presence of rotation, stratification and geometrical constraints \cite{young2017forward,byrne2013height,king2015upscale,siegelman2022moist}.

In the systems above, despite being close to two-dimensional,  variations along the third direction can not always be disregarded, and some times play an important role in the energy balance relation.
This is in particular true for the turbulent case where  
the dynamics of the flow drastically differ in two and in three dimensions.
Turbulence is realised both in three and in two dimensions when the Reynolds number $Re$ 
(the ratio of the viscous time scale to the to eddy turn-over time)
attains large values.
In three dimensional turbulence, the self-interaction between eddies
generates smaller and smaller eddies transferring energy
towards the smallest scales where it is dissipated effectively by viscosity,
independently of how small it is. 
%
In two dimensions, the opposite behavior is observed. 
Eddies self-organize to generate larger eddies and thus transfer energy to larger scales.
Unlike the three-dimensional case, 
in two dimensions, because energy is transferred upscale, viscosity is not efficient at dissipating it 
and energy piles up at the largest scale of the system.
The dynamics of flows in three and in two dimensions are thus fundamentally different.

This brings out the question: what happens when a flow is only approximately two-dimensional 
and the flow  lies between the two extreme situations? 
On the one hand,  if two dimensional motions dominate energetically one can expect that
flow dynamics will not be far from those of two dimensions and 
the transfer of energy will be towards the large scales.
On the other hand, the inverse transfer of energy can be proved only if {\it enstrophy}
(the mean square of vorticity)
is conserved exactly,
which is only true in two dimensional flows. 
Three dimensional perturbations, even if subdominant,  will break enstrophy conservation and the 
upscale transfer of energy is questioned. 
%
This question for the behavior of quasi-two dimensional flows has been around for a long time \cite{danilov2000quasi}.
What has been revealed in the last years is that a hybrid state of turbulence exists 
where transfer to larger and to smaller scales coexist, building what is referred to 
as a {\it bidirectional} or {\it split} cascade.

This review deals with when, how and in what sense can a flow transition from three dimensional to two dimensional behavior as a control parameter is varied. We will give a phenomenological description of the different {\it phases} of turbulence that are observed and describe their behavior close to the {\it critical points } where {\it phase transitions } are observed.
The richness of behaviors observed in quasi-two-dimensional turbulence surpassed all expectations leading to the discovery in the last years of new phenomena that challenge our mathematical and physical understanding of fluid turbulence. 
These new discoveries are reviewed and categorized in order to give direction to future research. 
Furthermore, the appearance of quasi-two-dimensional turbulence in a variety of different systems is also reviewed 
by mentioning key works in the different systems.
Finally, an attempt is made to note the many questions that are open in the field, stressing the need for further studies.  
Throughout the text, the mathematical formalism is made as light as possible in favor of readability and give references for more formal descriptions. 

The rest of this review is structured as follows. 
In section \ref{3D2D},  three and  two dimensional turbulence are briefly reviewed.
%
%
Then in section \ref{thin}, thin layer turbulence is discussed and  a detailed description of
the transition from three-dimensional (3D) to two-dimensional (2D) dynamics is given.
In section \ref{EXMPL2}, some recent results in other quasi-2D systems like rotating and magneto-hydrodynamic flows are given.
In section \ref{CON} conclusions are drawn and open questions in the field are presented.

\section{Two and three dimensional turbulence}  
\label{3D2D}                            
Turbulence in three and two dimensions has been a subject of study 
for well over a century, earning itself the title of the last unsolved problem of classical physics.
Some excellent books on three dimensional turbulence can be found in 
\cite{frisch1995turbulence,pope2000turbulent,davidson2015turbulence,davidson2011voyage} while 
reviews on three and two dimensional turbulence can be found in 
\cite{alexakis2018cascades, zhou2021turbulence, tabeling2002two, boffetta2012two, pandit2017overview}. 
The present review is limited in presenting  some basic results that are indispensable for the discussion that follows 
and refer the reader to the former mentioned reviews for any further information.

\subsection{Three dimensional turbulence}   
In its simplest form, turbulence in three and two dimensions  of a unit density fluid, is described by the evolution of the incompressible velocity field $\bf u$ that  follows the Navier-Stokes equation
\beq 
\partial_t \bu + \bu \cdot \nabla \bu = - \nabla P + \nu \nabla^2 \bu -\alpha \ou +  \bF
\label{NS3D}
\eeq 
where $P$ is the pressure imposing the incompressibility condition $\nabla\cdot \bu=0$,
$\nu$ is the viscosity and $\bF$ is an external force assumed to act on a lengthscale $\ell_f$
and inject energy at a rate $\cI_\cE$. The term  $-\alpha \ou$ (where $\ou$ stands for the vertically averaged velocity)
is an additional drag term that models the effect of boundaries and is 
added here to make contact later on with 2D turbulence.
We will consider as domain an orthogonal box of dimensions $L\times L \times H$ with $H$ being the height along the direction which fluctuations are suppressed.  
For simplicity, we consider here only periodic boundary conditions. 
For a given functional form of the forcing and non-dimensionalizing using $\ell_f$ and $\cI_\cE$
the resulting non-dimensional numbers of the system are 
(i) the Reynolds number is given by $Re=\frac{\cI_\cE^{1/3}\ell_f^{4/3}}{\nu}$
(ii) the large scale Reynolds number $Re_\alpha= \frac{\cI_\cE^{1/3}}{\ell_f^{2/3}\alpha}$
(iii) the length scale ratio $L/\ell_f$ and
(iv)  the normalized height $H/\ell_f$.
In what follows all quantities will be non dimensionalized using 
the forcing length-scale $\ell_f$ and the energy injection rate $\cI_\cE$ (thus setting  
$\ell_f=1$ and $\cI_\cE=1$). 

In three dimensions, the inviscid unforced system \ref{NS3D} for smooth flows, conserves two quadratic invariants. The first is energy $\cE=\frac{1}{2}\la \vert\bu\vert^2\ra$ where the angular brackets stand for volume average. The second is helicity $\cH=\frac{1}{2} \la \bu\cdot \bw \ra$ (where $\bw=\curl \bu$ is the vorticity) that  is not going to be discussed here but refer the reader to \cite{alexakis2018cascades,pouquet2022helical}.
In the presence of a forcing and dissipation the following energy balance relation holds 
\beq 
\cI_\cE = \epsilon_{\nu} + \epsilon_{\alpha} 
\label{balance3D}
\eeq 
where $\cI_\cE = \la \bu\cdot\bF \ra$ 
%
%
      is the time and volume averaged energy injection rate, 
      $\epsilon_{\nu} =\nu  \la \vert \nabla \bu {\vert} ^2\ra  $ is the energy dissipation rate due to viscosity and  
      $\epsilon_{\alpha} =\alpha  \la \vert \ou {\vert} ^2\ra  $ is the energy dissipation rate due to the drag term.

%
To express the notion of scale we are going to use the Fourier transformed fields $\hat{\bu}_\bk$ such that
\beq 
\hat{\bu}_\bk = 
\la e^{-i\bk\cdot \bx} \bu \ra \quad \mathrm \quad 
\bu = \sum_\bk e^{i\bk\cdot \bx} \hat{\bu}_\bk
\eeq 
where $\ell=1/k$ gives a natural definition of scale. Using this we can define the spherically averaged energy spectrum 
as 
\beq
E(k)= \frac{L}{2} \sum_{k\le \vert\bq\vert < k+1/L} \vert\hat{\bu}_\bq\vert^2
\eeq
where the sum is over all wavenumbers $\bq$ that satisfy $k\le \vert\bq\vert <k+1/L$. Note that we have normalized $E(k)$
by $L$ so that it has dimensions of energy per unit of wavenumber.

It was the pioneering work of last century that lead to the understanding that it is the flux of energy through scales that controls the statistical properties of high $Re$ flows. 
Early work of Kolmogorov \cite{kolmogorov1941local} argued for the existence of a constant flux of energy in scale space from the large to the small dissipative scales such that in the infinite $Re,Re_\alpha$ limit (in our notation)
\beq 
\epsilon_\nu = \cI_\cE, \quad \mathrm{and} \quad \epsilon_\alpha=0
\eeq 
so that all of the injected energy arrives at small scales and it is dissipated by viscosity.
Further assuming that this process is self-similar in scale, led to the prediction of the famous Kolmogorov energy spectrum 
\beq 
E(k) = c_{_K} \epsilon_\nu^{2/3} k^{-5/3}, \quad k_f<k<k_\nu
\eeq 
for wavenumbers larger than $k_f=1/\ell_f$ where $c_{_K}$ is a non-dimensional order one constant.
His prediction is valid up to what is now known as the Kolmogorov wavenumber
$k_\nu =1/\ell_\nu = \epsilon_\nu^{1/4}/\nu^{3/4}\simeq k_fRe^{-3/4}$.
For wavenumbers smaller than $k_f$ since there is no flux of energy these scales are expected to reach a {\it thermal equilibrium} state with an equipartition of energy among modes
leading to  
\beq 
E(k) = c_{_T} \cI_\cE^{2/3} \ell_f^{11/3}   \,\, k^2, \quad k<k_f
\eeq 
where $c_{_T}$ is an other non-dimensional constant \cite{dallas2015statistical,alexakis2019thermal,gorce2022statistical}.

Later research showed that the forward cascade is not self-similar and the distributions of velocity differences develop stronger tails as smaller scales are examined. As a result there is a small correction to the exponent of the energy spectrum.
This phenomenon referred to as {\it intermittency} has been the study of numerous studies 
in order to quantify and understand these corrections, that however is not going to be discussed here.  

\subsection{Two dimensional turbulence}    
\label{sec2D}                              

Despite the fact that two dimensional flows obey the same equation as three dimensional flows they do not display the same dynamical properties. In two dimensions the Navier-Stokes equation \ref{NS3D} can be written  in terms of the out-of-plane vorticity as 
\beq 
\partial_t w + \bu \cdot \nabla w = \nu \nabla^2 w  -\alpha w + f_w 
\label{NS2D}
\eeq 
where $w={\bf e_z}\cdot \bw $ (where ${\bf e_z}$ is the unit vector along  $z$  taken to be the out-of-plane direction),
$f_w = {\bf e_z}\cdot \curl \bF $ 
and we have added the drag force $-\alpha w$ often met in two dimensional systems due to bottom friction. 
There are two quadratic invariants in two dimensions the energy $\cE=\frac{1}{2}\la \vert \bu \vert ^2\ra$ and the enstrophy $\Omega=\frac{1}{2}\la \vert \bw \vert^2\ra$.
Their balance relations read
\beq 
\cI_\cE = \epsilon_{\nu} + \epsilon_\alpha, \quad 
\cI_\Omega = \eta_{\nu}  + \eta_\alpha  
\label{balance2D}
\eeq 
where 
$\cI_\cE$ is again the energy injection rate,
$\cI_\Omega = \la wf_w \ra \simeq k_f^2 \cI_\cE$ is the enstrophy injection rate.
The energy dissipation terms at small and large scales are given by
      $\epsilon_{\nu} = \nu \la   \vert \bw \vert^2 \ra$  and
      $\epsilon_{\alpha} = \alpha \la   \vert \bu \vert^2 \ra$ respectively
and the enstrophy dissipation rates by
      $\eta_{\nu} = \nu \la   \vert \nabla \bw \vert^2 \ra$  and
      $\eta_{\alpha} = \alpha \la   \vert \bw \vert^2 \ra$.
%

It was first realized by Onsager \cite{Onsager_1949} using a point vortex model that in two-dimensions negative ``temperature" states can exist where the flow self-organizes to generate large scale structures.
Later on, the work of Kraichnan, Leith and Batchelor 
\cite{kraichnan1967inertial,kraichnan1971inertial,leith1968diffusion,batchelor1969computation},
lead to what is known as the KLB dual cascade picture of two dimensional turbulence. 
Their work argued that a constant forward flux of energy is incompatible with 
a constant flux of enstrophy. In spectral space the energy spectrum $E(k)$ is related to 
the enstrophy spectrum $E_\Omega(k)=k^2E(k)$. This implies that a constant 
forward energy flux would imply an ever larger flux of enstrophy that is inconsistent with the enstrophy conservation. 
As a result, the only possibility is that enstrophy cascades {\it forward} towards the small scales while energy cascades {\it inversely} in the large scales. 
Therefore at infinite domains and in the infinite $Re,Re_\alpha$ limit
\beq
\epsilon_\alpha \simeq \cI_\cE, \quad \epsilon_\nu \simeq  0 
\quad \mathrm{and} \quad
\eta_\alpha \simeq 0, \quad \eta_\nu \simeq  \cI_\Omega. 
\eeq

The presence of this dual cascade picture has been verified in numerical simulations \cite{boffetta2010evidence,lilly1969numerical,lilly1972numerical}
and experiments \cite{byrne2011robust,von2011double,kelley2011spatiotemporal}.
Same arguments as in the three dimensional cascade 
lead to the prediction of an $E(k)\propto k^{-5/3}$ energy spectrum for the range of wavenumbers that an inverse cascade of energy is present $k_\alpha \ll k\ll k_f$ while the steeper spectrum $E(k)\propto k^{-3}$ is predicted for the range of wavenumbers $k_f\ll k\ll k_\nu$ 
that a forward enstrophy cascade is observed.
Here $k_\nu = \eta_\nu^{1/6}/\nu^{1/2} \simeq  k_f Re^{1/2}$ while $k_\alpha = \epsilon_\alpha^{1/2}/\alpha^{3/2} = k_f Re_\alpha^{-3/2} $. Note that the viscous wavenumber $k_\nu$ has a different scaling in 2D than 3D. We have also introduced a new cut-off wavenumber $k_\alpha$ (and lengthscale $L_\alpha=1/k_\alpha$) such that wavenumbers with $k<k_\alpha$ are severely damped by the drag force. 
The inverse cascade of energy is shown not to be intermittent \cite{boffetta2000inverse}
unless energy is injected in a fractal set \cite{sofiadis2023inducing}.

Unlike the the small scale viscous cut off $\ell_\nu=1/k_\nu$ that can become arbitrarily small as $Re\to \infty$ the large scale cut of $L_\alpha=1/k_\alpha$ is limited by the domain size $L$. Thus while for $L_\alpha \ll L$ the aforementioned inverse cascade and power laws  will 
be present, when $L\ll L_\alpha$ the inverse cascade will reach the domain size scale before the drag coefficient $\alpha$ becomes effective. 
In the latter case energy that arrives at the large scales will pile-up at the domain size in what is known as a spectral {\it condensate} \cite{Falkovich_1992}. In square periodic domains this condensate takes the form of two large counter rotating vortexes with large enough amplitude for the dissipation to balance the energy arriving from the small scales. 
This leads to the estimate for the energy of the condensate to be
\beq \cE \simeq \frac{\cI_\cE}{\alpha + \nu L^2}. \eeq 
This energy can be very large in particular in the $\alpha=0$ case that we will examine later on.
The cascade picture described before is then altered and steeper power-laws are observed.
%
%
The amplitude of the velocity fluctuations is so large that it brings the system to 
a quasi-equilibrium state that merits various equilibrium statistical approaches 
\cite{Onsager_1949,kraichnan1975statistical,robert1991statistical,
naso2010statistical,bouchet2012statistical,laurie2014universal,frishman2018turbulence,van2022geometric}.
Figure 1 shows two vorticity visualisations of two-dimensional flows such that on the left $L_\alpha < L$ while on the right $L_\alpha \gg L$. The condensate state has been realized in various experiments \cite{byrne2011robust,francois2013inverse}. The statistical properties of the inverse cascade case $L_\alpha\ll L$ and the condensate case $L_\alpha\gg L $ are so different that we are going to treat them separately for the quasi-2D case.

\begin{figure*}
\centering
\includegraphics[width=0.85\textwidth]{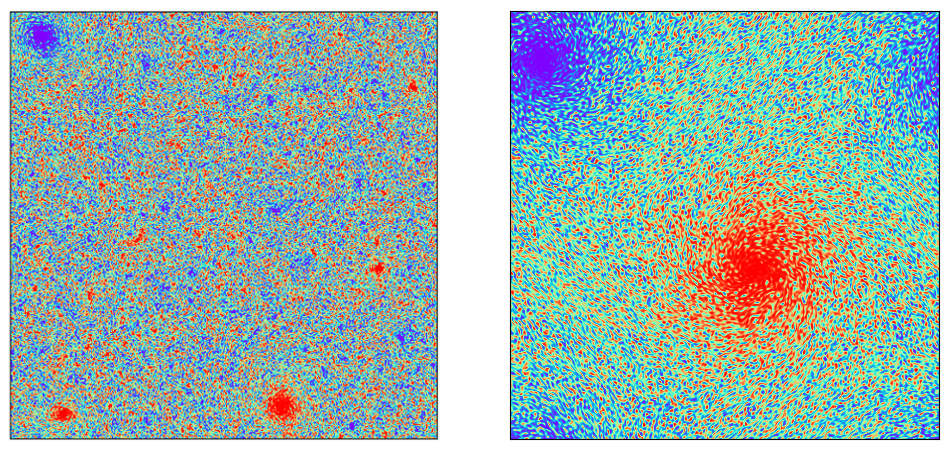}
\caption{Visualisation of vorticity for the turbulent state $L\gg L_\alpha$ (left) and for the condensate state $L_\alpha\gg L$ (right). In both cases the forcing scale $\ell_f$ was 80 times 
smaller than the domain size $L$.}
\label{fig1}
\end{figure*}

\section{Thin Layer turbulence} 
\label{thin} 

Turbulent flows confined in thin layers is perhaps the simplest and most intuitive flows that can display split energy cascades. It is thus an optimal choice for the study of quasi-two-dimensional flows. Furthermore, such anisotropic domains are very relevant for the atmosphere whose horizontal direction is of the order of 1000 km, while the pressure scale height of the order of 10 km. As such they have been extensively studied in the literature. 


Here we will discuss how the cascade properties of thin layer turbulence change as the layer height is varied in the limit of large $Re,Re_\alpha$. 
We will distinguish between two cases: 
(i) cases that the drag is efficient at absorbing the large scale energy that arrives in the presence of an inverse cascade
so that $L_\alpha\ll L$ and 
(ii) cases that the a large scale drag 
is very weak or absent so that a condensate is formed so that $L_\alpha \gg L$.
In the first case we will quantify the forward and inverse flux of energy 
using $\epsilon_\alpha$ and
$\epsilon_\nu$ (measured in units of $\cI_\cE$)
that express the fraction of energy that is transferred  in the large and the small scales respectively. 
In the infinite $Re,Re_\alpha$ limit $\epsilon_\alpha=0, \epsilon_\nu=1$ for three dimensional turbulence while $\epsilon_\alpha=1, \epsilon_\nu=0$ for two dimensional turbulence.

In the condensate case we will typically consider $\alpha=0$ so that $\epsilon_\alpha,\epsilon_\nu$ are not  suitable to quantify the state of the flow.
Instead we will use the energy of the largest scale modes $\cE_C$ defined as
\beq 
\cE_C =\frac{1}{2} \sum_{\vert \bk\vert < c/L} \vert \hat{\bu}_\bk \vert^2 
\eeq 
with $c>2\pi$ an order one number. 
Finally we will also consider the energy of $3D$ fluctuations as 
\beq 
\cE_{3D} =\frac{1}{2} 
\left\langle \vert \bu-\ou  \vert^2 \right\rangle.
\eeq 
In what follows we describe the behavior of $\epsilon_\alpha,\epsilon_\nu,\cE_C,\cE_{3D}$ as the height of the layer is varied for the turbulent and the condensate case.


\subsection{From 3D turbulence to 2D turbulence}                                    
%


First, the case $L_\alpha \ll L $ (that no condensate forms) is considered. 
The first work that made the remark that flows in thin layers can cascade energy to both large and small scales was \cite{smith1996crossover}. 
Since then a series of more systematic works followed
that measured the fraction of energy that cascaded to that large scales as the height was varied \cite{celani2010turbulence,musacchio2017split,benavides2017critical,van2019condensates,poujol2020role} whose results will be described in this section.

We begin the presentation of thin layer turbulence by considering first a layer of height $H$ much larger than the forcing scale
and then gradually decreasing it. For $H\gg \ell_f$ the flow displays three dimensional 
turbulence with only froward cascade observed at scales $\ell$ smaller than $\ell_f$ 
forming a $k^{-5/3}$ energy spectrum. At scales larger than $\ell_f$ energy is expected 
to reach a thermal equilibrium state with equipartition of energy among modes.
An argument can be made that if one 
considers horizontal scales $\ell_\perp \gg H \gg \ell_f$ where the flow is constrained 
to move primarily in two dimensions an inverse cascade can build up.  
However the primary interactions at such scales are not with same scale two dimensional eddies but rather 
directly with the forcing scale modes that are more energetic and act as a turbulent diffusion.
Thus even at these scales the cascade will be strictly forward. This is a conjecture however that needs to be verified. 

The strictly forward cascade behavior changes as smaller layer heights are considered.
Eventually a critical height $H_{3D}$ is reached such that for heights $H<H_{3D}$ a new phase of turbulence appears that
a bidirectional cascade is present. 
Numerical simulations indicate that the fraction of the energy that cascades inversely $\epsilon_\alpha$ is gradually increasing from zero as a power-law \cite{benavides2017critical,van2019condensates}
\beq 
\epsilon_\alpha \propto (H_{3D}-H)^{\beta_1} 
\eeq 
where $\beta_1$ is measured to be close to unity but its precise value
has yet to be determined.
The presence of this critical point is not trivial nor fully understood.
The only evidence we have are from numerical simulations \cite{benavides2017critical,van2019condensates} that 
however can be questioned because unavoidably they suffer from limited resolution. 
Further investigations would be required both numerical and theoretical to investigate
this point and conclude on its presence and on the universality class of this 
transition.

As $H$ is decreased further than $H_{3D}$ the fraction of energy that cascades
to the small scales is decreased. For $H \ll H_{3D}$ all scales $\ell>H$ 
have a two dimensional behavior with an inverse cascade of energy and 
a forward cascade of enstrophy. This however does not imply that no 
energy arrives at scales close to the height $\ell\sim H$. Along with the enstrophy cascade some energy 
has to be transported to the small scales as well, so at scale $H$ the flux of energy is $\epsilon_\nu \propto \eta_\nu H^2$ \cite{boffetta2011shell}. 
This energy is then transported to even smaller scales $\ell<H$ by three dimensional
interactions. Note that at the limit $H\to0$
the fraction of energy transported to the small scales also goes to zero $\epsilon_\nu\to 0$  
and 2D behavior is recovered. 
Given that $\eta_\nu \sim \cI_\Omega \simeq \cI_\cE/\ell_f^2$ implies that
\beq 
\epsilon_\nu \propto H^2
\label{H2}
\eeq
This behavior was first predicted in \cite{boffetta2011shell} and verified using a shell model.
This scaling however has never been tested using numerical simulations and it is something future 
research needs to confirm.

The scaling in \ref{H2} does not continue for arbitrarily small $H$. If $H\sim \ell_\nu$
then a new transition is observed towards a third phase of turbulence
where all three dimensional perturbations are damped out.
In particular in the case that the forcing is two-dimensional, the transition to 
exactly 2D behavior, occurs in a critical way: there is a second critical height 
$H_{2D}$ such that for all $H<H_{2D}$ the flow becomes exactly two dimensional. 
A very interesting dynamical behavior is observed for layer heights slightly
larger than $H_{2D}$. If we denote as $\cE_{3D}$ the energy contained in 3D modes alone
then it is measured that close to this new critical height $\cE_{3D}$ scales like
\beq 
\cE_{3D} \propto (H-H_{2D})^{\beta_2}
\eeq 
where $\beta_2$ is an exponent larger than one \cite{benavides2017critical,alexakis2021symmetry}.
This exponent is linked to the fact that as $H_{2D}$ is approached from above 
3D fluctuations grow or decay with growth that is randomly distributed in space and time.
As a result not only the amplitude of 3D perturbations depends on $H-H_{2D}$ but also the fraction of area that is occupied by them.
In \cite{alexakis2021symmetry}  this behavior was linked to the universality class observed in the presence of 
multiplicative noise in extended systems \cite{tu1997systems,genovese1998nonequilibrium}
that is related to the KPZ equation \cite{kardar1986dynamic}. 
This leads to the prediction that $\beta_2\simeq 1.7\dots$ compatible with present data
but further work is needed to draw firm conclusions.

\begin{figure*}[h]%
\centering
\includegraphics[width=0.85\textwidth,height=0.4\textwidth]{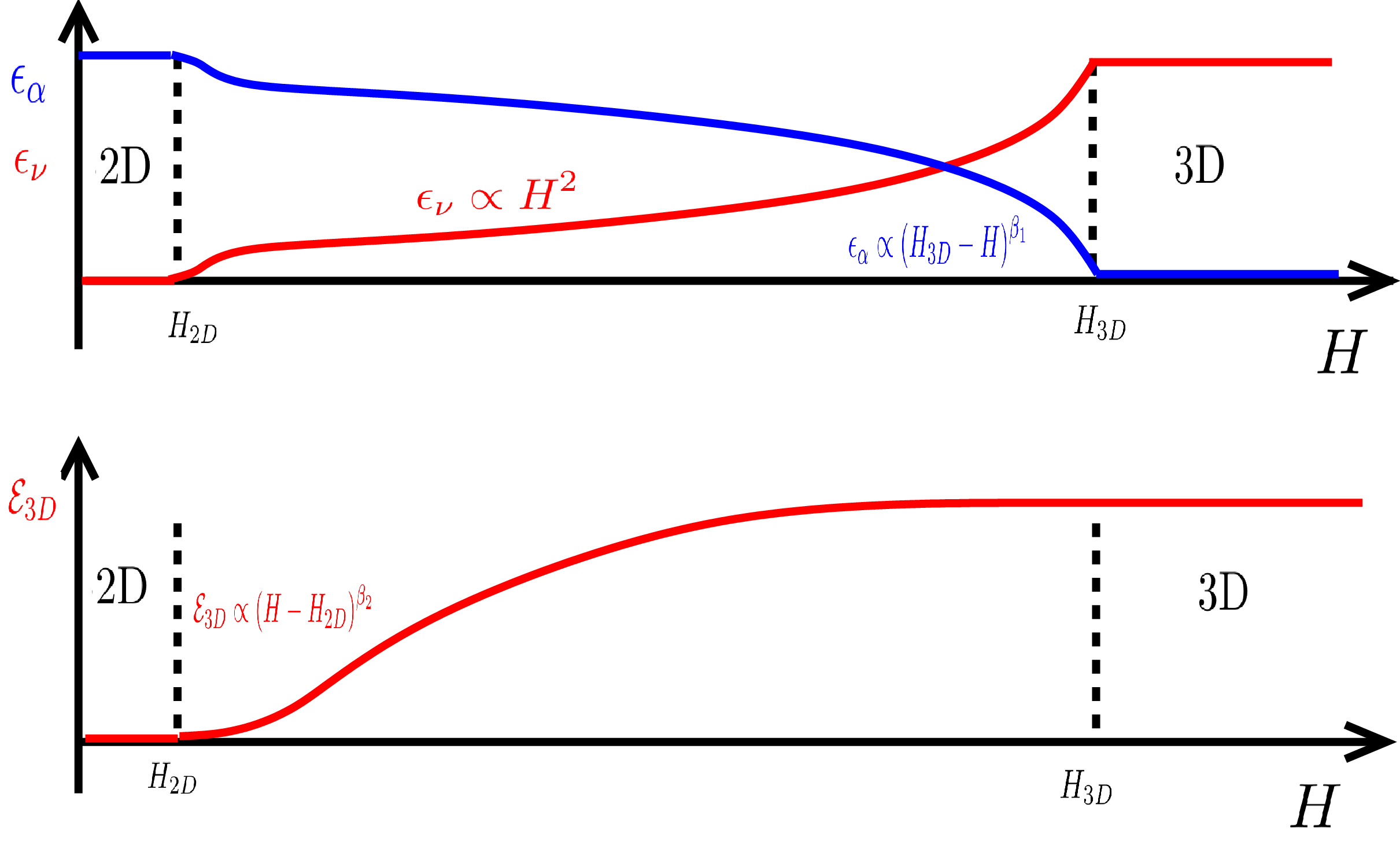}
\caption{ A sketch of the qualitative behavior of $\epsilon_\alpha$ (top panel, blue line), $\epsilon_\nu$ (top panel red line) and $\cE_{3D}$ (bottom panel red line) as a function of the layer height $H$.}
\label{fig2}
\end{figure*}

Finally for $H\ll\ell_\nu$ all three dimensional fluctuations are severely damped and the flow
can be proven to become exactly two dimensional \cite{gallet2015exact,gallet2016exact} and thus all properties mentioned in section \ref{sec2D} are recovered. 
A sum-up of all the different phases of thin layer turbulence and their transitions in the $L_\alpha \ll L$ case is sketched in fig \ref{fig2} and reported in table \ref{tbl}. 

\subsection{From 3D turbulence to a 2D condensate}                                  

In finite domains when the drag coefficient is very small (or even zero) 
even a weak inverse cascade will lead to the formation of a large scale condensate at long times.
The study of condensates is costly with numerical simulations due to the very long times that are involved. For this reason such studies always come after 
studies without the condensate. Nonetheless various numerical studies of thin layer turbulence 
exist in the literature \cite{van2019condensates,musacchio2019condensate}.
On the contrary to the numerical simulations, laboratory experiments do not have such time
limitations and various studies of thin layer flows have been investigated in the literature
although with limited separation between the forcing scale and the domain height \cite{xia2009spectrally,xia2011upscale,francois2013inverse,shats2010turbulence}. 
 
\begin{figure*}[h]%
\centering
\includegraphics[width=0.85\textwidth,height=0.4\textwidth]{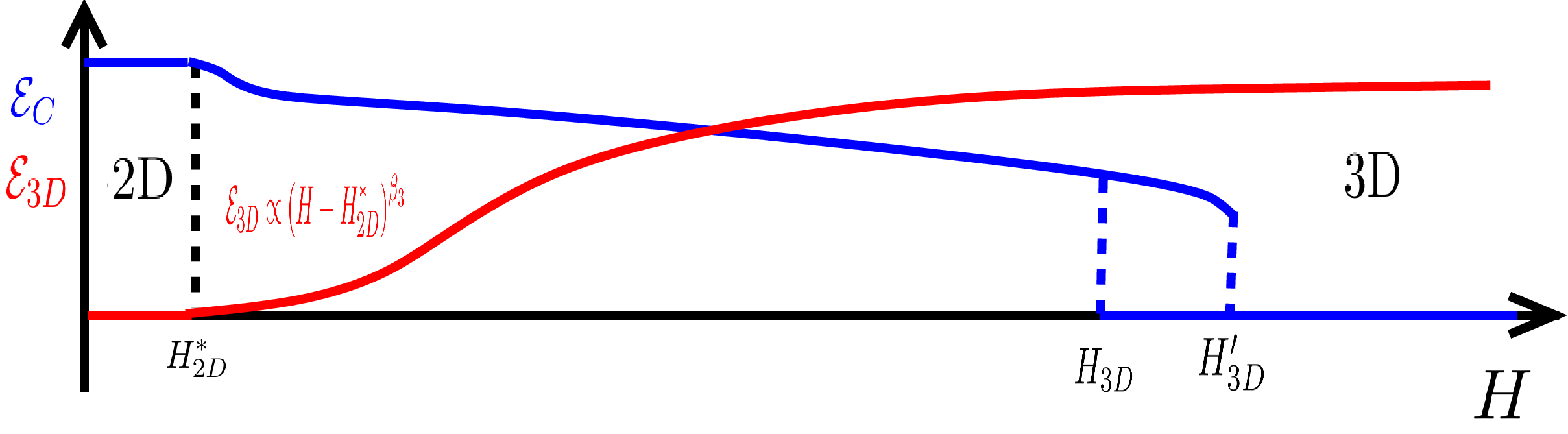}
\caption{ A sketch of the qualitative behavior of $\cE_C$ (blue line), and $\cE_{3D}$ (red line) as a function of the layer height $H$ in the condensate state.}
\label{fig3}
\end{figure*}

As in the previous section we begin with a layer that has a height $H$ much larger than the forcing scale $H\gg \ell_f$ so that no inverse cascade and no condensate is formed, and gradually reduce this height. 
As $H$ is reduced beyond the the critical height $H_{3D}$ (discussed in the previous section) a weak inverse cascade starts to build up leading to the formation of a condensate of energy $\cE_C$. If the drag coefficient $\alpha$ is finite the increase of  $\cE_C$ is gradual. For $\alpha=0$ however 
even a small inverse cascade can lead to a large value of $\cE_C$. In fact studies 
very close to $H_{3D}$ showed that the transition to the condensate state is discontinuous
\cite{van2019condensates}. In more detail it was shown that for $H$ larger than $H_{3D}$ the large scale energy remained small $\cE_C=\mathcal{O}(1/L^2)$,
but as soon as it become  slightly smaller than $H_{3D}$ a condensate formed and $\cE_C$ jumped to a finite value $\cE_C=\mathcal{O}(1)$. Furthermore, it was shown in \cite{van2019condensates} 
that if $H$ was gradually increased again to values larger than $H_{3D}$ the condensate state remained with $\cE_C=\mathcal{O}(1)$ up to some second critical value $H'_{3D}>H_{3D}$. 
Thus for values of $H$ in the range $H_{3D}<H< H'_{3D}$ two steady states (two different attractors) exist for the same value of $H$ and  a hysteresis diagram was constructed \cite{van2019condensates} as shown in fig.\ref{fig3}.

Further investigations of this system  \cite{van2019rare}  revealed that if run for a long time there are random jumps from one attractor to the other. The time distribution of these random jumps follows an exponential distribution indicating a memory-less process \cite{van2019rare}.
It was also shown that as the Reynolds number and the domain size are increased this range of $H$ where both attractors are stable increases \cite{de2022bistability}. This bistability is thus a behavior that is expected to survive in the large $Re$ limit.

As $H$ is decreased significantly from $H_{3D}$ so that $\ell_f\gg H$ the condensate becomes stable and its amplitude depends on the principal mechanism that saturates the inverse cascade.
If the Reynolds number is moderate, viscous dissipation will provide the main saturation mechanism and $\cE_C \propto \mathcal{I}_\cE L^2/\nu$ \cite{van2019condensates}.
For large Reynolds numbers however a new mechanism for saturation is present that originates from an eddy-viscosity effect due to the small 3D eddies at scales $\ell<H$. These 3D eddies extract energy from the condensate scales. A flux loop mechanism thus is present in which energy injected at $\ell_f$ moves up scale to the condensate scale $L$ through 2D motions and then back to smaller scales $\ell<H$ through interactions with the forcing scale 3D eddies. The condensate energy then in this case is not inversely proportional to viscosity but rather reaches a viscosity independent scaling $\cE_C\propto  (\mathcal{I}_\cE \ell_f)^{2/3}$ \cite{musacchio2019condensate,van2019condensates}. A simple model that captures this behavior was proposed in \cite{van2019condensates}. 

At even smaller layer heights, such that $H\sim  \ell_\nu$, a new transition appears again towards a state where 3D perturbations are damped as in the previous section. The case of the condensate however is significantly different from the turbulent case. While in the former 
turbulence was unstructured and uniform in space allowing 3D perturbations to grow anywhere in the domain, in the condensate state the flow
is still chaotic but self-organized in coherent structures of high concentration of vorticity and strain in small regions of space. In \cite{seshasayanan2020onset,lohani2023effect}
the evolution of infinitesimal 3D perturbations in a two dimensional flow in the condensate state were followed. This study revealed that 
 $\cE_{3D}$ in this limit followed a random behavior with long periods of decay and very short periods of very large exponential increase. The periods of increase
were shown to appear when the extreme of vorticity or the strain of a flow crossed a certain threshold. The statistics of these extremes were studied in detail recently in 
\cite{seshasayanan2023spatial}. Using a point vortex model that was coupled to point like 3D perturbations in \cite{van2021intermittency} showed that the instantaneous 
growth-rate of the energy of the 3D perturbations displayed power-law distributions that were linked to the power-law distribution of strain in space. 
As such the logarithm of the energy of the perturbations followed a L{\'e}vy random walk that explains the sudden jumps in the growth of energy observed in \cite{seshasayanan2020onset}.
This lead to a new type of intermittency the L{\'e}vy-On-Off intermittency that was described and studied in detain in \cite{van2021levy,van20231}. This also leads to new power-law behavior for the 3D energy 
\beq 
\cE_{3D} \propto (H-H_{2D}^*)^{\beta_3}  
\eeq 
where $H_{2D}^*$ is the critical value of $H$ bellow which all 3D perturbation are damped and
$\beta_3$ is an other exponent that depends on the L{\'e}vy noise parameters (see \cite{van2021levy,van20231}). This behavior suggested by the model has yet to be confirmed by direct numerical simulations.

Finally, for $H\ll\ell_\nu$ all 3D perturbations are dumped and the systems recovers its exact two dimensional behavior. The different stages of thin layer condensates are shown in figure \ref{fig3} and  summed up  in table \ref{tbl} along with the turbulent case. 



\begin{table}[h!] 
\centering
\begin{tabular}{|c||c|c|}
\hline
$ H \& L$    & Turbulent cascade     & Condensate  \\
             &   $L_\alpha\ll L$     &  $L_\alpha\gg L$ \\ \hline \hline
 $H>\ell_f$  & 3D Forward            &  No condensate     \\ 
             & Cascade of Energy     &                     \\
             & $\epsilon_\alpha = 0, \epsilon_\nu=1$        & $\cE_C \propto L^{-2}$  \\  \hline   
             & Criticality  & Discontinuous transition    \\
$H=H_{3D}\sim \ell_f$      &  $\epsilon_\alpha \propto (H_{3D}-H)^{\beta_1}$             & hysterisis and    \\ 
                &    $\beta_1\simeq1.$           & ``rare event" jumps    \\ \hline
                & 2D and 3D   & Flux-loop    \\ 
         $\ell_f \gg H \gg \ell_\nu$      & Bidirectional cascade  & condensate    \\ 
            & $\epsilon_\nu \propto H^2 $ &  $\cE_C \propto 1 $    \\ \hline
 $H \sim \ell_\nu$ & Extended Multiplicative noise  &  L{\'e}vy flight   \\ 
                &  $\cE_{3D}\propto (H-H_{2D})^{\beta_2} $   &   on-off Intermittency  \\ 
                &      $\beta_2 \sim 1.7$    & $\cE_{3D}\propto (H-H_{2D}^*)^{\beta_3} $ \\ \hline
$H\ll \ell_\nu$ & 2D Inverse  & 2D condensate  \\ 
              &  energy cascade    &     \\
             & $\epsilon_\alpha = 1, \epsilon_\nu=0$  &  $\cE_C \propto L^{2}/\nu$ \\ \hline
              
\end{tabular}
\vspace{0.3cm}
\caption{ Table with the different phases and critical behaviors observed in thin layer turbulence.
The left column gives the range of $H$, the middle gives the characteristics of the flow for the turbulent case and the right column gives  the characteristics of the flow for the condensate state.
\label{tbl}}
\end{table}
\section{Other examples of Quasi-2D flows}              
\label{EXMPL2}                        

Similar transitions to the ones observed in thin layer turbulence are 
expected to be found in other quasi-two-dimensional flows. However due
to the increased numerical cost or the technical difficulties in constructing laboratory
experiments has limited their study. In what follows we mention a few of these systems 
focusing on the additional phenomena that are present.  

\subsection{Rotating turbulence } 

Rotating turbulence 
is perhaps the simplest system after thin layers that displays quasi-2D behavior and split cascades.
It deals with turbulence in a rotating reference frame of rotation $\Omega$  quantified here by the Rossby number $Ro=\cI_\cE^{1/3} /(\Omega \ell_f)^{2/3}$. 
The transition from 3D turbulence to a 2D inverse cascade has been investigated in \cite{deusebio2014dimensional} and  \cite{pestana2019regime}. The critical layer height $H_{3D}$ was shown to increase with the rotation rate from the weak rotation value $H_{3D}\propto \ell_f$ to a value that was shown to increase as
\beq 
H_{3D} \propto \ell_f /Ro .
\eeq 
This scaling was confirmed in \cite{van2020critical} using an asymptotic model of the rotating Navier-Stokes equation for fast-rotating turbulence within highly elongated domains.

Although, rotating turbulence appeared to have the same phenomenological description as in thin layers for the 3D to 2D transition in  the large $Ro$ and small $Ro$ limit, at intermediate values of $Ro$ a new
state of turbulence was discovered in \cite{di2020phase} at which the flow formed a {\it crystal} of co-rotating vortexes  at scales larger than the forcing scale. This state of turbulence 
occurs close to the $H_{3D}$ transition height at which the co-rotating 2D vortexes are stable while counter-rotating 2D vortexes decay due to 3D fluctuations. The remaining 
co-rotating vortexes form crystals as have been seen in point vortex models \cite{aref1998point} and experiments on magnetized electron columns \cite{fine1995relaxation}.
This new state of turbulence is supported by a flux loop mechanism and was shown to be metastable that reduced to an inverse cascade if a strong perturbation is applied.
Such vortex crystals have been observed in the North pole of Jupiter by the Juno spacecraft mission \cite{adriani2018clusters} and have been interpreted by quasi-geostrophic dynamics 
in a curved domain \cite{siegelman2022polar}. The results of \cite{di2020phase} could provide an alternative explanation. We note that in both models a segregation between 
co-rotating and counter rotating  vortexes has lead to the formation of the crystal. 

Condensates in rotating turbulence have been studied in \cite{seshasayanan2018condensates,alexakis2015rotating,yokoyama2017hysteretic} 
where a discontinuous transition and hysteresis were also found close to $H_{3D}$, however the significant larger complexity of rotating turbulence did not allow for a thorough investigation. However, recently new experimental platforms have been build that have been able to quantify the formation of large scales in rotating turbulence and disentangle the forward and inverse transfers as well as the presence of wave turbulence \cite{lamriben2011direct,campagne2014direct,gallet2014scale,machicoane2016two, yarom2014experimental, kolvin2009energy,brunet2020shortcut,monsalve2020quantitative}. These experiments are very suitable in studying the long term behavior of rotating turbulence close to critical points and future work is expected to verify numerical observations as well as reveal new interesting physics.


\subsection{Stratified turbulence} 

By stratified we refer to turbulence in the presence of gravity $-{\bf e}_z g$ and a mean
stable density gradient $S=-\rho_0^{-1}d\rho/dz$.
Unlike rotation, stratification works against the two dimensionalization of turbulence by leading to the formation of strong vertical gradients, despite the fact that it suppresses vertical motions it. In  \cite{sozza2015dimensional} it was shown that the critical height $H_{3D}$ 
that the inverse cascade appears is decreased with stratification as  
\beq 
H_{3D} \propto \ell_f Fr,
\eeq
where $Fr=\cI_\cE^{1/3} /(gS)^{1/2}\ell_f^{2/3}$ is the Froude number. As a result 
in strongly stratified flows deviations from two dimensional turbulence can appear at much thinner layers than in the absence of stratification.  

Little or no work has been done for the other critical points in strongly stratified turbulence or in the presence of a condensate.


\subsection{Rotating and stratified turbulence} 

Rotating and stratified turbulence provides perhaps the simplest model of a dry atmosphere
and various studies have been devoted to its cascade properties directly from the Navier-Stokes  \cite{bartello1995geostrophic,marino2013inverse,pouquet2017dual,herbert2016waves}
or from reduced models \cite{xie2020downscale}.
It combines both the effect of two-dimensionalization
of rotation and the suppression of vertical motions of stratification.
The resulting complexity however is larger than the sum of its parts.
In \cite{van2022energy} using an asymptotic model for strong rotation
and long boxes it was shown that in this limit the parameter space 
split in three different phases, one with no inverse cascade, one
rotating dominated regime with an inverse cascade due to a two-dimensionalization
and a third strongly stratified regime with an inverse cascade corresponding
to quasi-geostrophic dynamics. It seems thus the phase space of 
rotating and stratified turbulence is far more complex than anticipated
and a careful exploration of it and its asymptotic limits is needed.

\subsection{Convection} 

Although convection involves inherently 3D motions in the presence 
of strong rotation or vertical confinement it can lead to quasi-2D behavior.
There have been numerous reports in the last years of an inverse cascade
and  the formation of coherent large scale quasi-2D vortices in rotating Rayleigh-Benard convection that coexists with 3D eddies that extract energy from the unstable stratification
\cite{guervilly2014large,favier2014inverse,maffei2021inverse}.
The behavior of the large scale vortices form resemble the ones observed in thin layer turbulence
and rotating turbulence displaying a discontinuous transitions and a hysteresis
\cite{favier2019subcritical,de2022discontinuous}.

Finally, inverse cascade has been reported in \cite{vieweg2022inverse} for very 
horizontally extended domains even in the absence of rotation.

\subsection{Magnetohydrodynamic turbulence} 

Like strong rotation for ordinary fluids a strong uniform magnetic field is also 
capable of bi-dimensionalizing electrically conducting fluids. Within the
Magneto-Hydro-Dynamic (MHD) approximation
two particular limits are of interest. First, the high conductivity limit (corresponding to the large magnetic Reynolds number
limit) relevant to astrophysical plasmas and tokamak devices. The transition to a two-dimensional behavior in this limit has been investigated in \cite{alexakis2011two,sujovolsky2016tridimensional} see also the review of  \cite{oughton2017reduced}. Second, for low magnetic Reynolds numbers (low conductivity)
the magnetic field acts as a damping mechanism for all velocity fluctuations varying along
the direction of the magnetic field. This so called quasi-static MHD
with relation to the transition to 2D flow has been studied in \cite{favier2011quasi,reddy2014strong,verma2017anisotropy}. This limit is suitable
for liquid metals and has been investigated in some innovating experiments \cite{potherat2014why,potherat2000effective,xia2011upscale,Xia_2003,gallet2012reversals}.

Just like in the previous examples in MHD transitions from a three dimensional forward cascade
to a bidirectional cascade and two dimensional cascade are observed. However the presence of
additional invariants both in 3D and 2D make the phase space of MHD turbulence sufficiently
more complex such that even in pure 2D MHD such transitions can be observed
\cite{seshasayanan2014edge,seshasayanan2016critical}.

\subsection{Quantum  turbulence} 
Quantum turbulence refers to turbulent flow of quantum fluids,  where vortex are quantized.  There are various models used to describe such flows that is beyond the purpose of this work to describe. We will however mention the experiments in \cite{seo2017observation,johnstone2019evolution,Gauthier_2019} where quasi-two dimensionalization in super-fluids is observed and the formation of large scale vortices.
Numerically such bi-dimensionalization has been observed in thin superfluid layers \cite{muller2020abrupt} using the Gross–Pitaevskii equation and in the two fluid description of superfluid turbulence in the presence of a counter-flow \cite{polanco2020counterflow}.

\section{Conclusions} 
\label{CON}           

If something is kept from this review, it should be the plethora of new dynamical 
phenomena that appear in quasi-2D turbulence when it is pushed to the right limits.
To begin with, even in the simplest case of a thin layer flow, different phases of turbulence are observed  in the infinite $Re$ limit. Unlike the commonly accepted expectations for homogeneous and isotropic turbulence for which a universal behavior
is expected, in the presence of confinement and anisotropy distinct phases of turbulence 
are present such that in one phase there is large scale
energy transfer and self-organization while in the other disorder and efficient energy dissipation. These phase are separated by critical points that display 
continuous or discontinuous phase transitions. Near these critical points, novel dynamical behaviors are observed that include the appearance of hysteresis diagrams and new critical exponents that are summed up in the table \ref{tbl}. As the complexity of the system is increased further new phases are discovered like the vortex-crystal meta-stable phase observed in rotating turbulence.

Recent research has only scratched the surface of these new phenomena and a lot
of further work is required.
In particular, investigating the behavior of the flows close to criticality and classifying them in universality classes is a much needed direction for the field. 
\NEW{Theoretical, numerical and experimental investigations need to proceed in parallel in this direction in order to establish  a clear and quantitative understanding of these transitions. }

Furthermore, expanding the system complexity including more physical effects would allow to make contact with with physical and industrial systems. \NEW{Here, only idealized situations were examined with a well defined injection and dissipation scale.  Reality is far more complex, with forcing mechanisms that span a wide range of scales (like convection, planetary scale baroclinic instabilities, etc) and large scale dissipation much more complex than the linear drag force assumed here. However, if we can not understand this behavior in the idealised models what hope do we have to understand the more complex physical systems. Progress in both directions and a connection between the the idealized and more physical case would be required in order to obtain accurate predicting models. }

Finally I would would like to note that, we live in a world that climate gradually changes and an atmosphere belongs in the wider class of quasi-two-dimensional flows. It is thus important to understand how this system responds to variations of parameters. For this reason studies of fundamental questions along the directions reviewed in this work are imperative. 

\


A peer-reviewed version of this article can be found at 
{\it Reviews of Modern Plasma Physics volume 7, Article number: 31 (2023)}
{\tt https://doi.org/10.1007/s41614-023-00134-3 }


\bibliographystyle{plain} 
\bibliography{Quasi2D_arxiv}    


\end{document}